\documentclass{article}

\begin{document}

\author{Abdullah Algin$^{1},$ Metin Arik$^{2},$ Ali S. Arikan$^{2}$ \and $^{1}$%
{\small Department of Physics, Osmangazi University, Me\c{s}elik, Eski\c{s}%
ehir, Turkey} \and $^{2}${\small Department of Physics,
Bo\u{g}azi\c{c}i University, Bebek, Istanbul, Turkey}}
\title{\textbf{Two-parameter deformed supersymmetric oscillators with SU}$%
_{q_{1}/q_{2}}$\textbf{(}$n\mid m$\textbf{)-covariance}}
\maketitle

\begin{abstract}
A two-parameter deformed superoscillator system with $SU_{q_{1}/q_{2}}$%
\textbf{(}$n\mid m$\textbf{)}-covariance is presented and used to
construct a two-parameter deformed $N=2$ SUSY algebra. The Fock space representation of the
algebra is discussed and the deformed Hamiltonian for such
generalized superoscillators is obtained.
\end{abstract}
\newpage

\section{Introduction}

A great deal of effort has recently been spent to the study of
many aspects of quantum groups and their associated algebras,
which are specific deformations of the usual Lie groups and Lie
algebras with some deformation parameter $q$\cite{a}-\cite{d}.
Many of such studies can be mentioned within a wide spectrum of
research of theoretical physics such as noncommutative geometry\cite{e},
two-dimensional conformal field theories\cite{f},
quantum mechanics\cite{g}.

After the realization of intimate relationship between
$q$-deformed bosonic
as well as fermionic oscillators and quantum groups (and their algebras)\cite{h}-\cite{j},
these relations have subsequently been
extended to two-parameter deformed versions of such oscillator
algebras\cite{k} and quantum group structures\cite{l}.
Meanwhile, several deformed forms of superalgebras
and supergroups have extensively been constructed by a natural
association of one or two-parameter deformed bosonic and fermionic
oscillator algebras\cite{m}-\cite{o}.

As is well known that the ordinary $N=2$ SUSY algebra developed by Witten\cite{p}
combines undeformed bosons with undeformed
fermions. This algebra has the following form:

\begin{equation}
\left\{ Q,Q^{\star }\right\} =H,\qquad Q^{2}=(Q^{\star })^{2}=0,\qquad \left[H,Q\right] =\left[ H,Q^{\star }\right] =0,
\end{equation}
where $Q$ and $Q^{\star }$ are odd generators called supercharges,
$H$ is an even generator called Hamiltonian. These generators are
assumed to be well defined on the relevant Hilbert space. After
this $N=2$ SUSY algebra has been constructed, several $q$-deformed
versions of this algebra have been proposed by using either
$q$-deformed boson operators or $q$-deformed fermionic operators\cite{r}-\cite{u}.
These studies have been done by mutually
commuting $q$-deformed bosonic and $q$-deformed fermionic
oscillator variables. Recently, such studies have also been extended to $q$%
-oscillator systems covariant under some quantum supergroup transformations\cite{m},\cite{v}.

The present paper investigates an interesting generalization for the $%
\noindent N=2$ SUSY algebra. In our generalization, the
superoscillators
system is accomplished by two independent real deformation parameters $%
(q_{1},q_{2})$ and has a covariance under the two-parameter
deformed quantum supergroup $SU_{q_{1}/q_{2}}(n\mid m).$ Moreover,
our generalization gives mutually non-commuting two-parameter
deformed bosons and fermions.

We should also mention that another example of such a mutually
non-commuting bosons and fermions property with $sl_{q}(n\mid
n)$-covariance was recently
introduced using the one-parameter deformed oscillator variables by Chung\cite{v}.
On the other hand, different algebraic forms
of SUSY structure such as fractional supersymmetry and
parasupersymmetry have been studied by taking deformation
parameter $q$ being a root of unity\cite{w}-\cite{y}.

In this paper, our aim is to construct a two-parameter deformed
$N=2$ SUSY
algebra by using the $SU_{q_{1}/q_{2}}(n\mid m)$-covariant $(q_{1},q_{2})$%
-deformed bosonic and $(q_{1},q_{2})$-deformed fermionic
oscillator system. However, our superoscillator algebra
construction serves as a generalization related to the studies on
the deformation of the conventional $N=2$ SUSY algebra.

The paper is organized as follows: In section 2, the two-parameter
deformed superoscillator algebra and its $SU_{q_{1}/q_{2}}(1\mid
1)$-covariance are shown. In section 3, we construct the
$(q_{1},q_{2})$-deformed SUSY quantum mechanics for
$SU_{q_{1}/q_{2}}(n\mid m)$-covariant $(q_{1},q_{2})$-deformed
boson and $(q_{1},q_{2})$-deformed fermion system. In section 4,
we analyze the Fock space representation of the
$(q_{1},q_{2})$-deformed $N=2$ SUSY algebra and obtain the
deformed Hamiltonian of the two-parameter deformed superoscillator
system. Finally, we give our conclusions in section 5.

\section{$SU_{q_{1}/q_{2}}(1\mid 1)$
-covariant two-parameter deformed superoscillator algebra}
In this section, we consider a system of one $(q_{1},q_{2})$ -deformed
bosonic and one $(q_{1},q_{2})$-deformed fermionic oscillators,
and show their covariance under the quantum supergroup $SU_{q_{1}/q_{2}}(1%
\mid 1)$. For this aim, we introduce the following
$(q_{1},q_{2})$-deformed superoscillator algebra:\bigskip
\begin{eqnarray}
AB &=&\frac{q_{1}}{q_{2}}BA, \\
AB^{\star } &=&q_{1}q_{2}B^{\star }A, \\
AA^{\star }-q_{1}^{2}A^{\star }A &=&q_{2}^{2(N_{b}+N_{f})}, \\
B^{2} &=&(B^{\star })^{2}=0, \\
BB^{\star }+q_{2}^{2}B^{\star }B
&=&q_{2}^{2(N_{b}+N_{f})}+(q_{1}^{2}-q_{2}^{2})A^{\star
}A=q_{1}^{2N_{b}}q_{2}^{2N_{f}},
\end{eqnarray}
where $A,\,A^{\star }$ and $B,B^{\star }$ are the deformed bosonic
and fermionic annihilation and creation operators, respectively.
By using the tensor product, the annihilation operators which
satisfy the above algebraic relations can be written as
\begin{eqnarray*}
A &=&a\otimes q_{2}^{N_{f}} \\
B &=&q_{1}^{N_{b}}\otimes c
\end{eqnarray*}
where $a$ is the deformed bosonic annihilation operator and $c$ is
the
deformed fermionic annihilation operator. $q_{1},q_{2}\in \mathbf{R}^{+}.$ $%
N_{b}$ and $N_{f}$ are the bosonic and fermionic number operators,
respectively. It is important to notice that, in this
consideration, the two-parameter deformed bosonic and fermionic
oscillators do not commute with each other (Eqs. (2) and (3)), and
the system still satisfies the Pauli exclusion principle (Eq.
(5)). Eq. (4) gives deformed bosonic commutation relation whereas
Eq. (6) gives the deformed fermionic anticommutation one.

By considering the limit $q_{2}=1$, one can easily realize that
above algebraic relations take the $SU_{q_{1}}(1\mid 1)$ covariant form\cite{m}.
It is helpful to remember this limiting case when one
tries to find the transformation matrix $T$ which leaves invariant
the system defined in Eqs. (2)-(6). In the light of these facts,
let us consider the following transformation\cite{v}:
\begin{equation}
\left(
\begin{array}{c}
A^{^{\prime }} \\
B^{^{\prime }}
\end{array}
\right) =T\,\left(
\begin{array}{c}
A \\
B
\end{array}
\right) =\left(
\begin{array}{cc}
a & \beta  \\
-(a^{\star })^{-1}\beta ^{\star }(a^{\star })^{-1} & (a^{\star
})^{-1}
\end{array}
\right) \left(
\begin{array}{c}
A \\
B
\end{array}
\right) ,
\end{equation}
where $T$ is the $2\times 2$ quantum super-matrix with two even
$(a,a^{\star })$ and two odd $(\beta ,\beta ^{\star })$
generators. Since these generators satisfy the following algebraic
relations:
\begin{eqnarray}
a\beta  &=&\frac{q_{1}}{q_{2}}\beta a,  \nonumber \\
a\beta ^{\star } &=&\frac{q_{1}}{q_{2}}\beta ^{\star }a, \\
\beta ^{2} &=&(\beta ^{\star })^{2}=0,\qquad \ \beta \beta ^{\star }+\frac{%
q_{1}^{2}}{q_{2}^{2}}\beta ^{\star }\beta =0,  \nonumber \\
aa^{\star }+\beta \beta ^{\star } &=&1,\qquad \ \ \ \ \ \ \ \ \ \
\ \ \ \ \ \ aa^{\star }-a^{\star }a=\left(
\frac{q_{1}^{2}}{q_{2}^{2}}-1\right) \beta ^{\star }\beta .
\nonumber
\end{eqnarray}
the matrix $T$ is an element of $ SU_{q_{1}/q_{2}}(1\mid 1).$
It is noticed that although the relations involving the oscillator
creation and
annihilation operators depend on the two deformation parameters $q_{1}$ and $%
q_{2}$, the relations containing the matrix elements of $T$
effectively involve a single parameter $r=q_{1}/q_{2}.$ This is
different from Celik's study\cite{z} where the quantum
matrix which leaves invariant another two-parameter
superoscillator algebra also depends on the two deformation
parameters separately.

According to Eq. (7), one can easily write the transformed
annihilation and creation operators as

\begin{eqnarray}
A^{^{\prime }} &=&a\otimes A+\beta \otimes B,  \nonumber \\
B^{^{\prime }} &=&-(a^{\star })^{-1}\beta ^{\star }(a^{\star
})^{-1}\otimes
A+(a^{\star })^{-1}\otimes B,  \nonumber \\
(A^{^{\prime }})^{\star } &=&a^{\star }\otimes A^{\star }-\beta
^{\star
}\otimes B^{\star }, \\
(B^{^{\prime }})^{\star } &=&-a^{-1}\beta a^{-1}\otimes A^{\star
}+a^{-1}\otimes B^{\star }.  \nonumber
\end{eqnarray}
In order to see all relations in Eqs. (2)-(6) remain unchanged for
transformed operators, we should also use braided tensor product
rule which can be shown as
\[
(x\otimes f_{1})(f_{2}\otimes y)=-xf_{2}\otimes f_{1}y\,,
\]
where $f_{1}$ and $f_{2}$ denote fermionic operators.

With the generalization of the above system to the
$(n+m)$-dimensional case, one can write the bosonic and fermionic
generators as
\begin{eqnarray*}
A_{i}
&=&\prod_{k=1}^{i-1}q_{1}^{(N_{b})_{k}}\,a_{i}\,%
\prod_{k=i+1}^{n}q_{2}^{(N_{b})_{k}}\,\prod_{k=1}^{m}q_{2}^{(N_{f})_{k}}, \\
B_{j}
&=&\prod_{k=1}^{n}q_{1}^{(N_{b})_{k}}\,%
\prod_{k=1}^{j-1}(-q_{1})^{(N_{f})_{k}}\,c_{j}\,%
\prod_{k=j+1}^{m}q_{2}^{(N_{f})_{k}},
\end{eqnarray*}
respectively. Here $i$ stands for $1$ to $n$ whereas $j$ stands
for $1$ to $m $. These generators satisfy the following algebraic
relations:
\begin{eqnarray}
A_{i}A_{j} &=&\frac{q_{1}}{q_{2}}A_{j}A_{i},\qquad \ \ i<j, \\
\qquad A_{i}A_{j}^{\star } &=&q_{1}q_{2}A_{j}^{\star }A_{i},\qquad
i\neq j,
\\
A_{i}A_{i}^{\star }-q_{1}^{2}A_{i}^{\star }A_{i}
&=&q_{2}^{2N}+(q_{1}^{2}-q_{2}^{2})\sum_{j=1}^{i-1}A_{j}^{\star }A_{j}, \\
A_{i}B_{k} &=&\frac{q_{1}}{q_{2}}B_{k}A_{i},\qquad \ \ \ \qquad  \\
A_{i}B_{k}^{\star } &=&q_{1}q_{2}B_{k}^{\star }A_{i},\qquad  \\
B_{k}B_{l} &=&-\frac{q_{1}}{q_{2}}B_{l}B_{k},\qquad \ \ \ k<l, \\
B_{k}B_{l}^{\star } &=&-q_{1}q_{2}B_{l}^{\star }B_{k},\qquad k\neq l, \\
B_{k}^{2} &=&(B_{k}^{\star })^{2}=0, \\
B_{k}B_{k}^{\star }+q_{2}^{2}B_{k}^{\star }B_{k}
&=&q_{2}^{2N}+(q_{1}^{2}-q_{2}^{2})\sum_{l=1}^{n}A_{l}^{\star
}A_{l}+(q_{1}^{2}-q_{2}^{2})\sum_{l=1}^{k-1}B_{l}^{\star }B_{l},
\end{eqnarray}
such that they are invariant under $SU_{q_{1}/q_{2}}(n\mid m)$
transformation. It is noted that in Eqs. (12) and (18), $N$
represents the total number operator which is nothing but the
addition of the fermionic and bosonic number operators. It is
clear that, in the limit $q_{2}=1,$ above system reduce the one
parameter deformed $SU_{q_{1}}(n\mid m)$-covariant superoscillator
algebra\cite{m}.

In the next section, the above deformed superoscillator algebra
will be used to construct a two-parameter deformed $N=2$ SUSY
algebra.

\section{ The construction of a two-parameter
deformed $N=2$ SUSY algebra for $SU_{q_{1}/q_{2}}(n\mid
m)$-covariant $(q_{1},q_{2})$-deformed bosons and fermions}

In this section, we construct the $(q_{1},q_{2})$-deformed SUSY
quantum mechanics for $n$ $(q_{1},q_{2})$-deformed bosonic and $m$ $%
(q_{1},q_{2})$-deformed fermionic oscillators covariant under the
quantum supergroup $SU_{q_{1}/q_{2}}(n\mid m).$ For the sake of
simplicity, we begin with the $n=m=1$ case. We have the following
deformed supercharges constructed from above deformed
superoscillators variables:
\begin{equation}
Q=A^{\star }\,B,\qquad Q^{\star }=B^{\star }A,
\end{equation}
where the operators $A,B$ satisfy the commutation
relations given in Eqs. (2)-(6). The odd generators $Q,Q^{\star }$
satisfy the nilpotency condition:
\begin{equation}
Q^{2}=(Q^{\star })^{2}=0,
\end{equation}
which can be obtained from Eq. (3), (5). From all considerations
above, we
have the following two-parameter deformed $N=2$ SUSY algebra with $%
SU_{q_{1}/q_{2}}(1\mid 1)$-covariance$:$%
\begin{equation}
\left\{ Q,Q^{\star }\right\} _{\frac{q_{2}^{4}}{q_{1}^{4}}} =\widetilde{H}%
=q_{1}^{-2}\left\{ q_{2}^{2N}\left[ A^{\star }A+\frac{q_{2}^{2}}{q_{1}^{2}}%
B^{\star }B\right] +(q_{1}^{2}-q_{2}^{2})(A^{\star
}A)^{2}\right\},
\end{equation}
\begin{eqnarray}
\left[ Q,\widetilde{H}\right] _{q_{2}^{4}/q_{1}^{4}} &=&0, \\
\left[ Q^{\star },\widetilde{H}\right] _{q_{1}^{4}/q_{2}^{4}} &=&0, \\
Q^{2} =(Q^{\star })^{2}&=&0,
\end{eqnarray}
where $N=N_{b}+N_{f},$ and also $\left\{ A,B\right\} _{r}=AB+rBA$
and $\left[ A,B\right] _{r}=AB-rBA.$ The deformed Hamiltonian
$\widetilde{H}$ in Eq. (21) gives a two-parameter generalization
of the Hamiltonian for the supersymmetric oscillator in quantum
mechanics\cite{ab}.

The algebra constructed in Eqs. (21)-(24) has some interesting
limiting cases: In the limit $q_{2}=1,$ we find the one-parameter
deformed $N=2$ SUSY algebra\cite{s},\cite{v}. The
conventional $N=2$ SUSY algebra in Eq. (1) can be recovered in the
limit $q_{1}=q_{2}=1.$ \

For arbitrary indices of deformed boson and fermion variables, we
have the following $2(n+m)$ supercharges:
\begin{equation}
Q_{i}=A_{i}^{\star }\,B_{i},\qquad Q_{i}^{\star }=B_{i}^{\star
}A_{i}.
\end{equation}
These supercharges are also nilpotent:
\begin{equation}
Q_{i}^{2}=(Q_{i}^{\star })^{2}=0,
\end{equation}
which can be obtained from Eqs. (14) and (17). Thus the
generalized
two-parameter deformed $N=2$ SUSY algebra for $SU_{q_{1}/q_{2}}(n\mid m)$%
-covariant $(q_{1},q_{2})$-deformed bosonic and
$(q_{1},q_{2})$-deformed fermionic oscillators is constructed by
the following deformed commutation and anti-commutation relations:
\begin{eqnarray}
\left\{ Q_{i},Q_{j}\right\} &=&0, \\
\left\{ Q_{i},Q_{j}^{\star }\right\} _{q_{2}^{2}/q_{1}^{2}} &=&0,
\end{eqnarray}
\begin{eqnarray}
\left\{ Q_{i},Q_{i}^{\star }\right\} _{\frac{q_{2}^{4}}{q_{1}^{4}}} &=&
\widetilde{H}_{i}=q_{1}^{-2}\left\{ q_{2}^{2N}\left[ A_{i}^{\star }A_{i}+(
\frac{q_{2}^{2}}{q_{1}^{2}})B_{i}^{\star }B_{i}\right]
+(q_{1}^{2}-q_{2}^{2})(\frac{q_{2}}{q_{1}})^{2}B_{i}^{\star
}B_{i}(\sum_{j=1}^{i-1}A_{j}^{\star }A_{j})\right\} \nonumber \\
&+&q_{1}^{-2}\left\{(q_{1}^{2}-q_{2}^{2})\left( A_{i}^{\star }A_{i}\right)
\left[\sum_{j=1}^{n}(A_{j}^{\star }A_{j})+\sum_{j=1}^{i-1}(B_{j}^{\star }B_{j})%
\right]\right\},
\end{eqnarray}
\begin{eqnarray}
\left[ Q_{j},\widetilde{H}_{i}\right] _{q_{2}^{2}/q_{1}^{2}} &=&0,\qquad \\
\left[ Q_{i},\widetilde{H}_{i}\right] _{q_{2}^{4}/q_{1}^{4}}
&=&0,\qquad
\end{eqnarray}
where $N=N_{b}+N_{f}.$ It is important to note that one can
recover the
one-parameter deformed $N=2$ SUSY algebra with $SU_{q_{1}}(n\mid m)$%
-covariance\cite{v} in the limit $q_{2}=1.$

\section{ Fock space representation of the
two-parameter deformed $N=2$ SUSY algebra with\\
$SU_{q_{1}/q_{2}}(n\mid m)$-covariance}

We now discuss the Fock space representation of the
two-parameter deformed $N=2$ SUSY algebra with
$SU_{q_{1}/q_{2}}(n\mid m)$-covariance. We first consider the
simplest case with $n=m=1.$ The bosonic and fermionic number
operators $N_{b}$ and $N_{f}$ satisfy the following commutation
relations:
\begin{eqnarray}
\left[ A,N_{b}\right] &=&A,\qquad \left[ A^{\star },N_{b}\right]
=-A^{\star
},  \nonumber \\
\left[ B,N_{f}\right] &=&B,\qquad \left[ B^{\star },N_{f}\right]
=-B^{\star }.
\end{eqnarray}
\bigskip We introduce the Fock basis $\left| n_{b},n_{f}\right\rangle $ and
the number operators also satisfy the following relations:
\begin{eqnarray}
N_{b}\left| n_{b},n_{f}\right\rangle &=&n_{b}\left|
n_{b},n_{f}\right\rangle
,\qquad n_{b}=0,1,2,...,  \nonumber \\
N_{f}\left| n_{b},n_{f}\right\rangle &=&n_{f}\left|
n_{b},n_{f}\right\rangle
,\qquad n_{f}=0,1, \\
N\left| n_{b},n_{f}\right\rangle &=&(n_{b}+n_{f})\left|
n_{b},n_{f}\right\rangle =n\left| n_{b},n_{f}\right\rangle ,
\nonumber
\end{eqnarray}
The representations of the operators $A,\,A^{\star
},\,B,\,B^{\star }$ are
\begin{eqnarray}
A\left| n_{b},n_{f}\right\rangle &=&q_{2}^{n_{f}}\sqrt{\left[ n_{b}\right] }%
\,\left| n_{b}-1,n_{f}\right\rangle , \\
\qquad \ \ \ A^{\star }\left| n_{b},n_{f}\right\rangle &=&q_{2}^{n_{f}}\sqrt{%
\left[ n_{b}+1\right] }\,\left| n_{b}+1,n_{f}\right\rangle , \\
B\left| n_{b},0\right\rangle &=&0,\qquad B\left|
n_{b},1\right\rangle
=q_{1}^{n_{b}}\,\left| n_{b},0\right\rangle , \\
\qquad B^{\star }\left| n_{b},1\right\rangle &=&0,\qquad B^{\star
}\left| n_{b},0\right\rangle =q_{1}^{n_{b}}\,\left|
n_{b},1\right\rangle ,
\end{eqnarray}
where
\begin{eqnarray}
A^{\star }A &=&\left[ N_{b}\right] \,q_{2}^{2N_{f}}=\left( \frac{%
q_{2}^{2N_{b}}-q_{1}^{2N_{b}}}{q_{2}^{2}-q_{1}^{2}}\right) \,q_{2}^{2N_{f}}\,, \\
B^{\star }B &=&N_{f}\,\,q_{1}^{2N_{b}},
\end{eqnarray}
which can be deduced from Eqs. (4) and (6). Acting the
supercharges on the Fock basis $\left| n_{b},n_{f}\right\rangle ,$
we have
\begin{eqnarray}
Q\left| n_{b},0\right\rangle &=&0,\qquad \ \ \ \ \ \ Q^{\star
}\left|
n_{b},1\right\rangle =0, \\
Q\left| n_{b},1\right\rangle &=&q_{1}^{n_{b}}\sqrt{\left[ n_{b}+1\right] }%
\,\left| n_{b}+1,0\right\rangle , \\
Q^{\star }\left| n_{b},0\right\rangle
&=&q_{1}^{n_{b}-1}\sqrt{\left[ n_{b}\right] }\,\left|
n_{b}-1,1\right\rangle .
\end{eqnarray}
The energy eigenvalues for the deformed Hamiltonian in Eq. (21)
are
\begin{eqnarray}
H\,\left| n_{b},1\right\rangle &=&\left(
\frac{q_{2}}{q_{1}}\right)
^{4}q_{1}^{2n_{b}}\left[ n_{b}+1\right] \,\left| n_{b},1\right\rangle , \\
H\,\left| n_{b},0\right\rangle &=&q_{1}^{2(n_{b}-1)}\left[
n_{b}\right] \,\left| n_{b},0\right\rangle .
\end{eqnarray}

We now discuss a generic case for the Fock space representation of
the
two-parameter deformed superoscillator algebra generators $%
A_{i},\,A_{i}^{\star }$ and $B_{i},\,B_{i}^{\star }.$ We introduce
the Fock
basis $\left| \left\{ \widetilde{n}_{b}\right\} ,\left\{ \widetilde{n}%
_{f}\right\} \right\rangle $ as follows:
\begin{eqnarray}
(N_{b})_{i}\,\left| \left\{ \widetilde{n}_{b}\right\} ,\left\{ \widetilde{n}%
_{f}\right\} \right\rangle &=&(n_{b})_{i}\,\left| \left\{ \widetilde{n}%
_{b}\right\} ,\left\{ \widetilde{n}_{f}\right\} \right\rangle
,\qquad
(n_{b})_{i}=0,1,2,..., \\
(N_{f})_{i}\,\left| \left\{ \widetilde{n}_{b}\right\} ,\left\{ \widetilde{n}%
_{f}\right\} \right\rangle &=&(n_{f})_{i}\,\left| \left\{ \widetilde{n}%
_{b}\right\} ,\left\{ \widetilde{n}_{f}\right\} \right\rangle
,\qquad
(n_{f})_{i}=0,1, \\
N_{i}\,\left| \left\{ \widetilde{n}_{b}\right\} ,\left\{ \widetilde{n}%
_{f}\right\} \right\rangle &=&(N_{b}+N_{f})_{i}\,\left| \left\{ \widetilde{n}%
_{b}\right\} ,\left\{ \widetilde{n}_{f}\right\} \right\rangle
=n_{i}\,\left| \left\{ \widetilde{n}_{b}\right\} ,\left\{
\widetilde{n}_{f}\right\} \right\rangle ,
\end{eqnarray}
where we have used the abbreviation for the Fock basis

\noindent $\left| \left\{ \widetilde{n}_{b}\right\} ,\left\{ \widetilde{n}%
_{f}\right\} \right\rangle =\left|
(n_{b})_{1},(n_{b})_{2},....,(n_{b})_{n};(%
\,n_{f})_{1},(n_{f})_{2},...,(n_{f})_{m}\right\rangle .$ The
representations of the operators \noindent $A_{i},\,A_{i}^{\star
}$ are
\begin{eqnarray*}
A_{i}\,\left| \left\{ \widetilde{n}_{b}\right\} ,\left\{ \widetilde{n}%
_{f}\right\} \right\rangle
&=&q_{1}{}^{\sum_{k=1}^{i-1}(n_{b})_{k}}q_{2}{}^{%
\sum_{k=i+1}^{n}(n_{b})_{k}}q_{2}{}^{\sum_{k=1}^{m}(n_{f})_{k}}\sqrt{\left[
(n_b)_{i}\right] }\,\left| \left\{ \widetilde{n}_{b}\right\}
-1,\left\{
\widetilde{n}_{f}\right\} \right\rangle \\
A_{i}^{\star }\,\left| \left\{ \widetilde{n}_{b}\right\} ,\left\{ \widetilde{%
n}_{f}\right\} \right\rangle
&=&q_{1}{}^{\sum_{k=1}^{i-1}(n_{b})_{k}}q_{2}{}^{%
\sum_{k=i+1}^{n}(n_{b})_{k}}q_{2}{}^{\sum_{k=1}^{m}(n_{f})_{k}}\sqrt{\left[
(n_b)_{i}+1\right] }\,\left| \left\{ \widetilde{n}_{b}\right\}
+1,\left\{ \widetilde{n}_{f}\right\} \right\rangle ,
\end{eqnarray*}
where
\begin{eqnarray}
A_{i}^{\star }A_{i}
&=&\,(q_{1}^{2})^{\sum_{k=1}^{i-1}(n_{b})_{k}}\,\left[
(n_b)_{i}\right] \,(q_{2}^{2})^{\sum_{k=i+1}^{n}(n_{b})_{k}}\,(q_{2}^{2})^{%
\sum_{k=1}^{m}(n_{f})_{k}},\phantom{AA} \\
\left| \left\{ \widetilde{n}_{b}\right\} \mp 1,\left\{ \widetilde{n}%
_{f}\right\} \right\rangle &=&\left|
(n_{b})_{1},...,(n_{b})_{i}\mp
1,...,(n_{b})_{n};(\,n_{f})_{1},...,(n_{f})_{m}\right\rangle ,
\nonumber
\end{eqnarray}
and $\left[ (n_b)_{i}\right] $ is defined by Eq. (38).

For the fermionic sector, we have the following number operator
for a generic case:
\begin{equation}
B_{i}^{\star
}B_{i}=(n_{f})_{i}\,(q_{1}^{2})^{\sum_{k=1}^{n}(n_{b})_{k}}\,(q_{2}^{2})^{%
\sum_{k=i+1}^{m}(n_{f})_{k}}\,(-q_{1})^{\sum_{k=1}^{i-1}2(n_{f})_{k}},
\end{equation}
which can be deduced from the defining relation of $B_{i}$ in
section 2. Therefore, the representations of the operators
$B_{i},\,B_{i}^{\star }$ are
\begin{equation}
B_{i}\,\left| \left\{ \widetilde{n}_{b}\right\} ,\left\{ \widetilde{n}%
_{f}\right\} \right\rangle
=\left\{
\begin{array}{l}
0\qquad\qquad \qquad \qquad\qquad \qquad\qquad \qquad \qquad    \mbox{if}\quad
(n_{f})_{i}=0, \\
q_{1}{}^{\sum_{k=1}^{n}(n_{b})_{k}}\,q_{2}{}^{\sum_{k=i+1}^{m}(n_{f})_{k}}%
\,(-q_{1}){}^{\sum_{k=1}^{i-1}(n_{f})_{k}}\,\left| \left\{ \widetilde{n}%
_{b}\right\} ,\left\{ \widetilde{n}_{f}\right\} -1\right\rangle  \\
\phantom{0}\qquad\qquad \qquad\qquad\qquad \qquad\qquad \qquad
\qquad     \mbox{if}\quad (n_{f})_{i}=1,
\end{array}
\right.
\end{equation}

\begin{equation}
B_{i}^{\star }\,\left| \left\{ \widetilde{n}_{b}\right\} ,\left\{ \widetilde{%
n}_{f}\right\} \right\rangle =\left\{
\begin{array}{l}
0\qquad\qquad \qquad\qquad\qquad\qquad \qquad\qquad \qquad
\mbox{if}\quad
(n_{f})_{i}=1,\\
q_{1}{}^{\sum_{k=1}^{n}(n_{b})_{k}}\,q_{2}{}^{\sum_{k=i+1}^{m}(n_{f})_{k}}%
\,(-q_{1}){}^{\sum_{k=1}^{i-1}(n_{f})_{k}}\,\left| \left\{ \widetilde{n}%
_{b}\right\} ,\left\{ \widetilde{n}_{f}\right\} +1\right\rangle   \\
\phantom{0}\qquad\qquad\qquad \qquad\qquad\qquad \qquad\qquad \qquad \mbox{if}
\quad(n_{f})_{i}=0, \\
\end{array}
\right.
\end{equation}
where $\,\left| \left\{ \widetilde{n}_{b}\right\} ,\left\{ \widetilde{n}%
_{f}\right\} \mp 1\right\rangle =\left|
(n_{b})_{1},...,(n_{b})_{n};(\,n_{f})_{1},...,(n_{f})_{i}\mp
1,...,(n_{f})_{m}\right\rangle .$

\section{Conclusions}
In this paper, we defined a two-parameter deformed
superoscillator algebra with $SU_{q_{1}/q_{2}}(n\mid
m)$-covariance. By means of such
generalized superoscillator system, we constructed a two-parameter deformed $%
N=2$ SUSY algebra covariant under the quantum supergroup $%
SU_{q_{1}/q_{2}}(n\mid m).$ We explicitly studied for the case of one $%
(q_{1},q_{2})$-deformed boson and one $(q_{1},q_{2})$-deformed
fermion system with $SU_{q_{1}/q_{2}}(1\mid 1)$-covariance. For
this system, we particularly discussed the Fock space properties
and found the energy eigenvalues for the deformed Hamiltonian in
terms of two deformation parameters. The two-parameter deformed
$N=2$ SUSY algebra constructed here has some important limiting
cases: The one-parameter deformed $N=2$ SUSY algebra\cite{v}
can be recovered in the limit $q_{2}=1.$ The limit
$q_{1}=q_{2}=q$ gives the $SU(n\mid m)$-covariant one-parameter
deformed $N=2$ SUSY algebra constructed from the $q$-deformed
bosonic and fermionic Newton oscillators\cite{u}. The
conventional $N=2$ SUSY algebra in Eq. (1) can be obtained in the
limit $q_{1}=q_{2}=1.$ \

This work was supported by Bogazici University Research Fund,
project number $02$B$301$.

\end{document}